\documentclass{emulateapj}
\usepackage{natbib}

\slugcomment{To Appear in ApJ.}
\shorttitle{SN 1961V Survived}
\shortauthors{Van Dyk \& Matheson}

\begin{document}

\title{It's Alive! The Supernova Impostor 1961V\footnote{Based on
observations made with the NASA/ESA {\sl Hubble Space Telescope}, obtained in
part from the data archive of the Space Telescope Science Institute, which is
operated by the Association of Universities for Research in Astronomy, Inc.,
under NASA contract NAS 5-26555.}}

\author{Schuyler D.~Van Dyk}
\affil{Spitzer Science Center, Caltech, 220-6, Pasadena, CA 91125}

\and

\author{Thomas Matheson}
\affil{National Optical Astronomy Observatory, 950 North Cherry
  Avenue, Tucson, AZ 85719}

\begin{abstract}
  Reports of the death of the precursor of Supernova (SN) 1961V in NGC
  1058 are exaggerated.  Consideration of the best astrometric data shows 
  that the star, known as ``Object 7,'' lies at the greatest proximity to SN 1961V 
  and is the likely survivor of the ``SN impostor'' super-outburst. 
  SN 1961V does not coincide with a neighboring radio source and is therefore not a radio SN. 
  Additionally, the current properties of Object 7, 
  based on data obtained with the {\sl Hubble Space Telescope},  
  are consistent with it being a quiescent Luminous Blue Variable (LBV). 
  Furthermore, post-explosion non-detections by the {\sl Spitzer Space Telescope\/} 
  do not necessarily and sufficiently rule out a surviving LBV. 
  We therefore consider, based on
  the available evidence, that it is yet a bit premature to reclassify 
  SN 1961V as a {\it bona fide\/} SN. The inevitable
  demise of this star, though, may not be too far off.
\end{abstract}


\section{Introduction}

The evolution of the most massive stars is not well known. It is thought that main sequence
stars with $M_{\rm ZAMS} \gtrsim 30\ M_{\sun}$ proceed through a blue supergiant phase,
possibly to a red supergiant phase, or straight to a luminous blue variable (LBV) phase, and
to the Wolf-Rayet (WR) phase prior to explosion as supernovae (SNe). That WR stars typically
possess $M\lesssim 20\ M_{\sun}$ \citep{crowther07} requires their precursor stars to shed
most of their mass, presumably through eruptive mass ejections as a LBV 
\citep{humphreys94,smith06}.
The term ``supernova impostor'' has been coined \citep{vandyk00} to describe these eruptive 
events for massive stars, such as $\eta$~Carinae \citep[e.g.,][]{davidson97}. This is because 
various observational aspects of the eruptions can mimic the properties of true SNe.
During $\eta$~Car's Great Eruption in the 1800s,
the star greatly exceeded the Eddington limit, with the
bolometric luminosity increasing by $\sim$ 2 mag.  
The total luminous output of such an eruption ($\sim 10^{49}$--$10^{50}$ erg) can rival that
of a SN.

However, a SN, by strict definition, is an explosive event that ends the life of a star 
(although a compact object may form in the process). If an optical transient is observed with
energetics comparable to that of a true SN,
and, after a sufficient period of time has elapsed, a star still exists at the exact position of
the transient, then that transient is a SN impostor.

The most studied example of a SN impostor, SN~1961V in NGC 1058,
remains controversial to this day. A debate has continued as to
whether or not this event was a true SN.  In fact, it is truly
impressive how many journal pages have been spent on this one object.
This includes two recent studies, both of which present arguments that
claim SN 1961V as a genuine SN. We argue here 
that SN 1961V is
an impostor and that its precursor has survived what was a powerful
eruption. Many of our points below were first presented in our review
of SN impostors \citep{vandyk11}.

\section{Supernova or Supernova Impostor?}

\begin{figure}
\figurenum{1}
\plotone{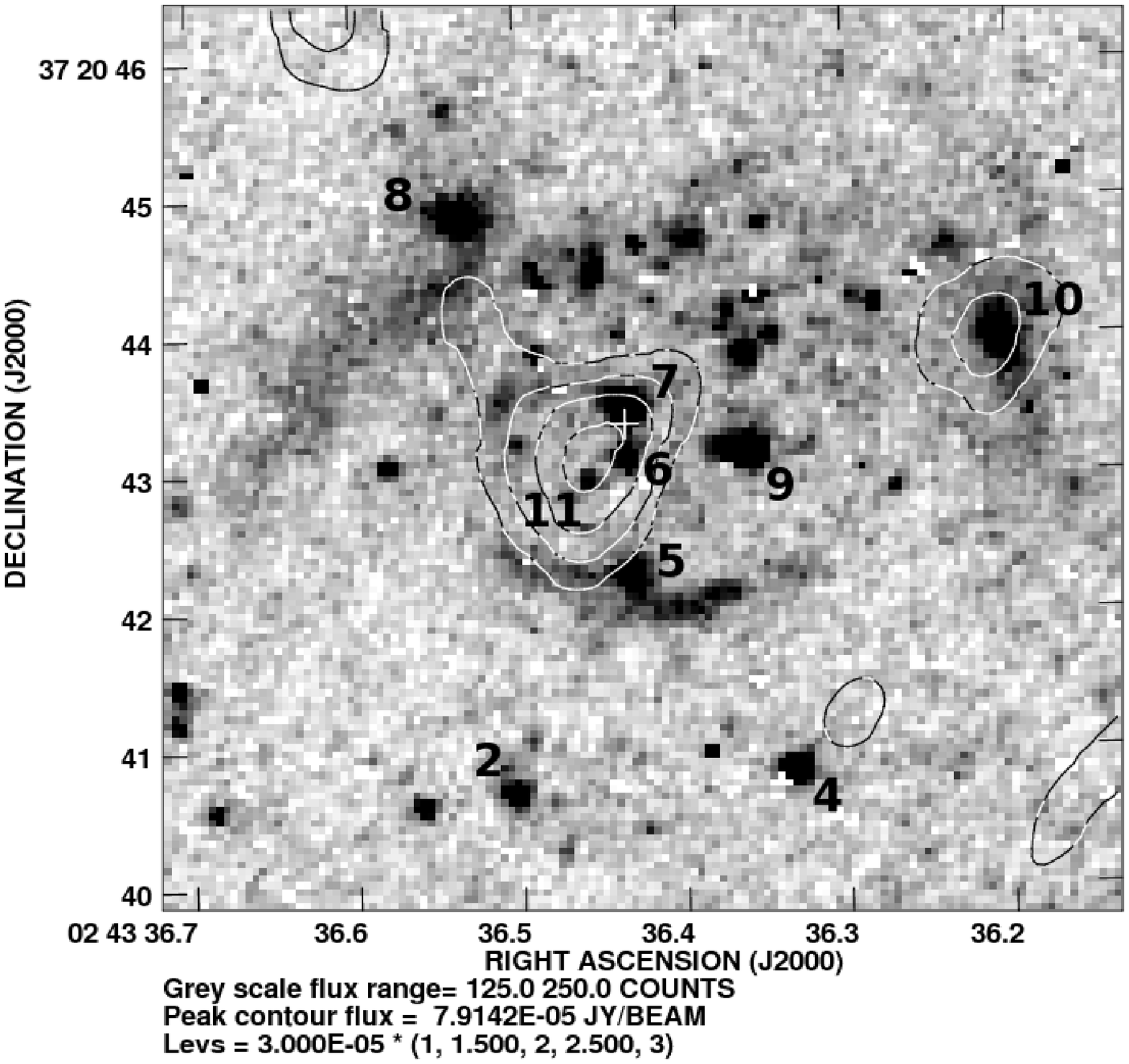}
\caption{{\sl HST\/}/STIS unfiltered 50CCD image of the SN 1961V site from \citet{chu04}, 
obtained from the archive.
Source numbering is the convention from \citet{filippenko95} and \citet{vandyk02}.  
The {\it contours\/} are the 6 cm radio data from \citet{stockdale01}, which we have
reanalyzed.  Contour levels are (1, 1.5, 2, 2.5, 3)$\times$0.03 mJy/beam.
The {\it white cross\/} represents the accurate absolute optical position of SN 1961V from 
\citet{klemola86} and is shown at $2\times$ the actual positional uncertainty.\label{radiomap}}
\end{figure}

We only briefly summarize the history of observations of SN 1961V and the 
debate. We encourage the reader to peruse the original
papers for more detailed discussions, which include (but are not limited to)
\citet{goodrich89}, \citet{humphreys94}, \citet{filippenko95}, \citet{humphreys99},
\citet{chu04}, \citet{vandyk05a}, \citet{smith11}, and Kochanek, Szczygiel, \& Stanek (2011).
While reading this summary, please keep in mind that the most 
accurate astrometry inevitably shows that the source ``Object 7'' \citep{filippenko95} is 
most coincident 
with SN 1961V, and that the centroid of detected radio emission is offset from this position. 
See Figure~\ref{radiomap}.

The photographic light curve for SN 1961V has been characterized as ``strange''
\citep{bertola64,branch71}. The progression of the outburst is described by
\citet[][their Table 3]{humphreys94} and shown schematically by, e.g., \citet{goodrich89}.
Additionally, the long-term photometric monitoring \citep{bertola65} of SN 1961V showed that its 
color remained relatively constant, 
abnormally resembling that of an F supergiant ($B-V \approx 0.6$ mag).
The SN faded below optical detectability in late 1968.

\bibpunct[]{(}{)}{;}{a}{}{;}
The unusual early-time optical spectra were dominated by
narrow emission lines of H, He\,{\sc i}, and Fe\,{\sc ii}, with a maximum expansion velocity of 
only $\sim 2000$ km s$^{-1}$ \citep{zwicky64,branch71}. These spectral 
attributes could be considered similar to the characteristics of Type IIn SNe 
\citep[SNe~IIn, where ``n'' designates ``narrow''][]{schlegel90, filippenko97}. 
For both impostors and SNe IIn the coincidence of a 
hydrogen envelope with similar kinetic energies results in spectra with similar appearance, 
specifically, strong Balmer lines,
typically a narrow emission profile (with velocities $\sim$100--200 $\,$km$~$s$^{-1}$)
atop an ``intermediate width'' base 
\citep[with velocities of $\sim$1500--2000 $\,$km$~$s$^{-1}$ in the case of
the SNe IIn;][]{smith08c}. The narrow component is interpreted as emission from pre-shock 
circumstellar gas. 
The intermediate-width component for SNe~IIn arises from interaction of the 
SN shock with the dense circumstellar matter \citep[e.g.,][]{turatto93,smith08c}, 
which is expected to result in luminous and long-lived radio, X-ray, and
optical emission 
\citep[e.g.,][; although, see \citeauthor{vandyk96}~\citeyear{vandyk96} 
regarding the radio emission]{chevalier94, chevalier03}.
See a recent discussion of SNe IIn and their properties by \citet{kiewe12}.
For SN impostors, the intermediate-width component of the Balmer emission lines 
are significantly narrower than for the SNe IIn, and early-time spectra for SNe IIn generally
are quite blue, whereas impostors, soon after discovery, can appear either ``hot'' or
``cool'' \citep{smith11}.

Both \citet{bertola64} 
and \citet{zwicky64} identified a $m_{\rm pg} \approx 18$ mag object at the SN site in 
pre-SN photographic plates of the host galaxy.  \citeauthor{zwicky64} noted that the star 
exhibited small variations for 23 years prior to rising by six magnitudes in a year, ``acting
similarly to $\eta$ Carinae....''  At the host galaxy's distance 
\citep[adopting 9.3 Mpc, the Cepheid distance to NGC~925, in the same ``NGC~1023'' group as NGC~1058;][]{silbermann96}, 
the presumed precursor 
star would have been an incredible $M_{\rm pg} \approx -12$ mag, implying it was the 
most luminous known star in the local Universe.  From this \citet{utrobin84} modelled the SN's 
peculiar light curve as the explosion of an (almost certainly improbable) 2000 $M_{\odot}$ star.

The hunt for the remnant of this extraordinary explosion began in earnest in 1983, 
when \citet{fesen85} identified a faint knot of H$\alpha$ emission very near the SN
position, in the easternmost of two H\,{\sc ii} regions 
\citep[each separated by $\sim$4\arcsec, with $\log L_{{\rm H}\alpha} \approx 37.2$;][]{goodrich89}.
\citet{branchcowan85}  
detected a source of nonthermal radio emission in 1984 with the Very Large Array (VLA), also 
very near the optical SN position.  \citet{klemola86} measured a very accurate position 
(uncertainty $\lesssim 0{\farcs}1$) for 
SN 1961V from Lick plates taken in 1961, as well as of the precursor from a plate from 1937.
The conclusion that \citeauthor{branchcowan85} reached was  
that the radio emission was from an aging SN, or very young remnant, corresponding to 
SN~1961V.   However, in very late-time optical spectra from 1986 of the 
SN environment, 
\citet{goodrich89} found a narrow (unresolved) H$\alpha$ emission-line profile atop 
an intermediate-width ($\approx 2100$ km s$^{-1}$ FWHM) base, at the SN position.
From a comparison with the properties of known LBVs, 
\citeauthor{goodrich89}~concluded that SN~1961V was {\it not\/} a true SN, 
but instead was a possible 
``$\eta$~Car analog.'' Comparing the H$\alpha$ emission they detected from SN 1961V to that
from $\eta$~Car, they argued that the survivor of a 1961 super-outburst lies behind
behind high circumstellar extinction of $A_V \approx 5$ mag.

\citet{filippenko95} subsequently obtained $VRI$ images of the SN field in 1991 
with the pre-refurbishment {\sl Hubble Space Telescope\/} ({\sl HST}) 
Wide-Field/Planetary Camera-1 (WF/PC-1),
detecting an apparent cluster of 10 stellar-like objects, all within $\sim$5\arcsec.  
From those relatively low-quality data, they isolated ``Object~6'' in the cluster
(see Figure~\ref{radiomap}) as the likely  
survivor of the 1961 super-outburst, although other stars in the
environment could not be completely excluded based on 
the positions of these stars as measured by \citeauthor{filippenko95}
Object 6, with $V \approx 25.6$ mag, has colors consistent with a red supergiant, 
possibly as a result of substantial dust ($A_V \approx 4$ mag).

If SN~1961V were the giant eruption of an $\eta$~Car analog, this still could not account for the 
nonthermal radio emission; $\eta$~Car itself, for example, is a thermal bremsstrahlung radio
source  \citep[e.g.,][]{cox95}.
From subsequent VLA observations in 1999 and 2000
\citet{stockdale01} confirmed the presence of nonthermal radio emission.
Furthermore, \citeauthor{stockdale01}~claimed that the source had declined in flux since the 
previous radio observations \citep{branchcowan85}. They inferred, once again, that SN~1961V is 
a fading, core-collapse radio SN.  If the radio source were the SN, then \citet{vandyk02}
identified in shallow, archival {\sl HST\/} Wide Field Planetary Camera 2 (WFPC2) 
images from 1994 and 2001 
a fainter, red, possible optical counterpart in the environment, ``Object~11.''

\citet{chu04} obtained  {\sl HST\/} Space Telescope Imaging Spectrograph (STIS) 
G430M and G750M grating 
spectra, along with an unfiltered 50CCD image, of the SN environment in 2002. 
They expected to detect emission lines, such as [O\,{\sc iii}]~$\lambda$5007, 
with high velocities, $>2000$ km s$^{-1}$, from an old SN, e.g., as seen from SN~1957D
\citep[e.g.,][]{long92}. Furthermore, if there were an old core-collapse SN in the field, 
detectable optical emission should
accompany the nonthermal radio emission, as is the case generally for SNe 
\citep[e.g.,][]{chevalier94}. 
\citeauthor{chu04} instead detected exactly {\it one\/} bright emission line, i.e.,
H$\alpha$, from exactly {\it one\/} point source within the 2{\arcsec}-wide slit.
They measured a line width for the broad profile of $\lesssim$550 km s$^{-1}$ full width zero
intensity. 
To an astrometric accuracy of ${\sim}0{\farcs}25$ using the STIS data, Chu et al.~attributed the 
spectrum to Object~7, 
due to its coincidence with the SN optical position 
\citep{klemola86}. They compared the H$\alpha$ line to that of $\eta$~Car and
concluded that this star is a possible LBV in the field. However, given the presence of the 
nonthermal radio emission, assumedly associated with an old SN 1961V, \citeauthor{chu04} 
did not directly associate Object~7 with SN~1961V, except as a possible binary companion
to the SN's progenitor.

\citet{vandyk05a} and \citet{vandyk11} 
summarized and countered several of these arguments above, continuing
to maintain the impostor status for SN 1961V. However,
recently, two studies, \citet{smith11} and \citet{kochanek11}, 
have concluded that SN 1961V was not an impostor, but, instead, a {\it bona fide\/}
SN. \citeauthor{smith11}~base their conclusion on the relative energetics of 
the event and its apparent overall similarity to the SNe IIn, specifically the narrow emission-line
widths and slow and unusual light curve. These authors argue that the high peak luminosity and
high expansion velocity make it problematic for SN 1961V to conform to other presumed 
SN impostors and
known LBVs. They argue that the light curve is consistent with the superposition of a normal
Type II-Plateau SN light curve and a SN IIn-like light curve, with 
additional luminosity arising from enhanced interaction at late times
with an existing circumstellar medium. 
\citeauthor{kochanek11} rely almost entirely on an analysis of the mid-infrared
dust emission in the SN environment. Their hypothesis is that, 
since the prediction is that a LBV survivor of SN 1961V would be cloaked by 4--5 magnitudes of 
visual extinction, this dust shell should be easily detectable in the thermal infrared (IR) as an IR
$\eta$~Car analog.
The lack of detectable mid-IR emission at the SN position led \citeauthor{kochanek11} to 
conclude that SN 1961V must have been a true SN, which has now faded below optical 
detectability.

\section{Observations of SN 1961V}

There have been multiple observations of SN 1961V at a wide range of
wavelengths in the last fifty years.  Depending on the assumptions
used in interpreting these data sets, one can come to several
plausible conclusions.  Here, and in \S~\ref{analysis}, we make the case that 
the overall picture across the electromagnetic spectrum is simpler and more consistent with SN 
1961V as an eruptive LBV.

\subsection{Radio Data}

\bibpunct[ ]{(}{)}{;}{a}{}{;}
We obtained from the VLA data archive the 6 cm data, originally
collected by \citet{stockdale01}, and produced a map using standard
routines in NRAO's Astronomical Image Processing System (AIPS). We
obtained from the {\sl HST\/} archive the STIS 50CCD image from
\citet{chu04} and applied a very accurate astrometric grid, $\pm
0{\farcs}02$ in both right ascension and declination, to the image,
using the tasks {\it geomap\/} and {\it geotran\/} in
IRAF\footnote{IRAF (Image Reduction and Analysis Facility) is
  distributed by the National Optical Astronomy Observatory, which is
  operated by the Association of Universities for Research in
  Astronomy, Inc., under cooperative agreement with the National
  Science Foundation.}. We then registered this image to our version
of the radio map. This analysis first appeared in \citet{vandyk05a}.

The radio emission overlaid on the {\sl HST\/} image is shown in
Figure~\ref{radiomap}.  The objects from \citet{filippenko95} and
\citet{vandyk02} are labeled in the figure. We place the accurate
optical absolute position of SN 1961V \citep{klemola86} in the figure
as well.  It is clear from the figure that the optical SN position is
most coincident with Object~7 \citep[][also pointed out this positional
coincidence]{chu04}.  However, the centroid of the radio source, which
both \citet{cowan88} and \citet{stockdale01} claim to be associated
with the SN, is significantly displaced, by $\sim 0{\farcs}4$
southeast (i.e., $\gtrsim 4\sigma$), from the optical position.

\bibpunct[; ]{(}{)}{;}{a}{}{;}
Furthermore, the 
nonthermal radio emission does not appear that well correlated with the stellar objects in the 
environment.  Instead, we consider it far more likely that it is 
emission associated with what appears to be a complex, turbulent interstellar 
medium. Note the ``finger'' of radio emission that extends to the northeast from the main central 
concentration of emission, which appears to follow the arcs 
of nebular emission seen in the STIS image toward the central cluster of stars.
The unfiltered STIS image \citep{chu04}, which clearly includes nebular emission lines,
alludes to such a spectacular interstellar 
environment (which is very conspicuous as being the most optically luminous 
extranuclear emission region in the host galaxy).  This is the ``eastern H\,{\sc ii} region'' from
\citet[][also \citeauthor{cowan88} \citeyear{cowan88}]{fesen85} seen at much higher spatial resolution.

Additionally, our flux density measurement from the 2000 radio 6 cm
data is 0.091$\pm$0.005 mJy.  Comparing this to 0.11$\pm$0.03 mJy from
1986 \citep{cowan88}, it is not entirely obvious that the radio flux density
had, in fact, actually declined in that time interval, as
\citet{stockdale01} claimed. Therefore, we consider it to be quite
convincing that the radio emission is not that of a fading radio SN, nor does
the radio emission have any direct relation at all to SN 1961V.

\subsection{{\sl HST\/} Optical Data}

We have analyzed archival {\sl HST\/}/WFPC2 images, specifically of SN
1961V, obtained in 2008 August by program GO-10877 (PI: W.~Li). These are
exposures of 360 sec total in F555W and 800 sec total in F658N (the
{\sl HST\/} narrow band that includes H$\alpha$ at approximately the
host galaxy redshift). The images are shown in
Figure~\ref{hstimage}. The stars that are visible in the figure
from \citet{filippenko95} and \citet{vandyk02}, again, are
labeled. Admittedly not of the highest signal-to-noise ratio ($S/N$), it is
clear from the figure that Object 7 is well detected in both bands (at
$5.7\sigma$ in both F555W and F658N).  In fact, Object 7 is the
solitary H$\alpha$ source in the immediate central cluster. Object 10
(which from Figure~\ref{radiomap} is most likely an emission region) and
possibly another, unnumbered object to the northeast of Object 7, are
the only other detected point-like sources in that image.  

We measured photometry for these images using the routine HSTphot
\citep{dolphin00a,dolphin00b}, which automatically accounts for WFPC2
point-spread function (PSF) variations and charge-transfer effects
across the chips, zeropoints, aperture corrections, etc. HSTphot was
run on both bands with a 3$\sigma$ detection threshold. Object 7 is
detected at $m_{\rm F555W}=24.70 \pm 0.19$ and $m_{\rm F658N} = 20.45
\pm 0.19$ mag.

\bibpunct[]{(}{)}{;}{a}{}{;}
We can compare the $m_{\rm F555W}$ measurement for 
Object 7 from these 2008 data with that from the WF/PC-1 observations
by \citet{filippenko95}, i.e., $m_{\rm F555W}=24.22 \pm 0.07$ mag,
and the WFPC2 F606W measurement by \citet{vandyk02}, i.e., 
$m_{\rm F606W}=23.84 \pm 0.14$ mag. Furthermore, we can measure 
the brightness of Object 7 from the archival STIS 50CCD image 
(normalizing to a 24-pixel-radius, or $1{\farcs}2$-radius, aperture, which 
includes 100\% of the encircled energy, following the online STIS Instrument 
Handbook), adopting the appropriate zero point for the clear (CL) bandpass from 
\citet{rejkuba00}, and find $m_{\rm CL}=24.41 \pm 0.04$ (which is $\sim$V, if we
assume $V-I\simeq 0.5$ mag).
Although it is likely that we cannot
directly compare observations made with {\sl HST\/} with different
detectors and different bandpasses to better than $\sim$10\%,
the $m_{\rm F555W}$ and $m_{\rm CL}$ all agree relatively well, while the
$m_{\rm F606W}$ differs pretty significantly from the others. 
However, we have checked the photometry from \citet{vandyk02} by
running the images they analyzed through the most recent version of
HSTphot and obtain $m_{\rm F606W}=24.44 \pm 0.19$ mag. 
This is quite
different from the original, published value, but is far more
in line with both of the $m_{\rm F555W}$ magnitudes.  
\citep[We cannot
readily explain why HSTphot now produces a significantly different
result than found by][.]{vandyk02}
All-in-all, the $\sim V$ magnitudes for Object 7 are consistent, although
we cannot complete rule out some variability or gradual fading 
of the source over nearly the last two decades.

We also obtained the STIS spectral data \citep{chu04} from the {\sl HST\/} archive and 
re-extracted the spectrum using standard STSDAS routines within IRAF. 
The portion of the spectrum including H$\alpha$ is shown in Figure~\ref{figspec}.
Although the spectrum is noisy, both a broad and a narrow component to the line are
quite evident.

\subsection{{\sl Spitzer\/} Mid-Infrared Data}\label{spitzer}

Similarly to \citet{kochanek11}, we analyzed the archival mid-IR data for NGC 1058,
obtained using the {\sl Spitzer Space Telescope\/} with both the IR Array Camera (IRAC; 3.6,
4.5, 5.8, and 8.0 $\mu$m) and the Multiband Band Photometer for {\sl Spitzer\/} (MIPS; we
analyzed the 24 $\mu$m data only). We considered the 
observations using both instruments from 2004 (GTO program 69; PI: G.~Fazio) 
and from 2007 (GO program 40619; PI: R.~Kotak), 
and assuming no variability for SN 1961V between these two epochs (if actually 
detected), we combined the data from these observations for each of the bands.
The data we analyzed corresponded to pipeline versions S18.7 for IRAC and S18.12 for MIPS.
We used the MOPEX \citep[MOsaicking and Point source EXtraction;][]{makovoz05a,makovoz05b} 
package provided by the {\sl Spitzer\/} Science Center to mosaic the 
individual Basic Calibrated Data (BCDs; in fact, for IRAC we used the artifact-corrected CBCDs) to
produce a single image mosaic in each band (for both IRAC and MIPS, we left the first frame out
of each set of observations when mosaicking, since it often has a far shorter exposure time than
the rest of the BCDs and therefore adds mostly noise to the mosaic). We also applied the array 
location-dependent photometric corrections to the IRAC CBCDs within MOPEX, although, given 
the number of CBCDs and the adequate redundant coverage, this correction was not particularly
important in the end. Although emission in all bands is 
detected from the environment of SN 1961V, the emission in the resulting mosaics is diffuse.
As \citet{kochanek11} point out, neither Object 7 nor 
any of its immediate neighboring sources are detected in any of the {\sl Spitzer\/} bands. Object 8,
which is well separated from the SN 1961V position, dominates the emission from the 
environment 
at 8.0 $\mu$. At 24 $\mu$m the spatial resolution is too poor to resolve which source, or sources,
is the primary emitter in the environment. There is little point in analyzing the relatively 
low-resolution, low-sensitivity 70 $\mu$m data for the host galaxy.

We then used the routine APEX (Astronomical Point source 
EXtractor) Single Frame, with the ``user list'' input option, within MOPEX to perform aperture 
photometry at the exact position of SN 1961V. For the IRAC mosaics we used a 3.0-pixel-radius
aperture, with an annulus for sky subtraction of radius 12.0--20.0 pixels, computing the sky 
background using the mode within the annulus. For the MIPS 24 $\mu$m mosaic we employed
a 1.22-pixel-radius aperture, with sky annulus of radius 8.16--13.06 pixels. We applied the
aperture corrections for the IRAC bands for our aperture/annulus configuration from the online
IRAC Instrument Handbook\footnote{http://ssc.spitzer.caltech.edu/irac/iracinstrumenthandbook/, 
Table 4.7.}). To determine the correction for the MIPS aperture, we had also performed
point response function (PRF) fitting photometry on the mosaic. We considered the PRF fluxes of
the two brightest stars, seen in the mosaic well away from the body of the galaxy, as ``truth.''
We then computed the ratio of the fluxes measured for these two stars through our 
aperture/annulus configuration and the PRF fluxes, i.e., 4.0:1.0, and corrected the aperture flux
at the SN 1961V position by this ratio. All of these upper limits to the detection of SN 
1961V are shown in Figure~\ref{figmir}. (Since these are only upper limits, we dispensed with 
applying color corrections to both the IRAC and MIPS photometry.) Our limits are comparable to, 
although generally higher than, those that \citet{kochanek11} have estimated. The pertinent 
values with which to compare are those in their Tables 2 and 3, labelled ``SN1961V area'' 
(particularly, their $3{\farcs}6$-radius aperture measurements for IRAC), i.e., 
$<$0.026, $<$0.022, $<$0.079, $<$0.230, and $<$0.265 mJy (ours) {\it versus\/} 
$<$0.023, $<$0.016, $<$0.069, $<$0.207, and $<$0.226 mJy (theirs), 
at 3.6, 4.5, 5.8, 8.0, and 24 $\mu$m, respectively.

\section{Analysis}\label{analysis}

\bibpunct[; ]{(}{)}{;}{a}{}{;}
One of the assumptions made by \citet{kochanek11}, as well as by \citet[][and by earlier 
authors]{smith11}, is that the object detected in photographic plates back to the 
1930s, prior to the 1960 ``S Dor-type eruption'' \citep{humphreys94}
and the 1961 luminous event was, in fact, the quiescent progenitor. We are now 
increasingly skeptical of this, and, instead, we find it far more compelling to presume that
the precursor star was already in a state of sustained outburst prior to the more
energetic eruption in the 1960s \citep[see also][]{goodrich89,humphreys99}. 
Clearly, this is difficult, if not impossible, to prove, due to the 
nonexistence of observations prior to the first available plates. 
However, the $B-V$ color ($\simeq 0.6$ mag)
measured by \citet{utrobin87} from the 1954 Palomar Sky Survey plates is essentially the same as 
the color during the outburst in 1961 and 1962 \citep{bertola65}, and is  
the expected color of a LBV in an ``eruptive state'' \citep{humphreys94}. 
Furthermore, when the object was in its pre-outburst state at $m_{\rm pg} \approx m_B \simeq 18$ 
mag, even assuming only Galactic foreground extinction \citep[$A_V=0.2$ mag;][]{schlegel98}, 
it had $M_{\rm bol} \approx -12.6$ mag, which is well above the modified 
Eddington limit, $M_{\rm bol} \simeq -11$ mag \citep[e.g.,][]{ulmer98}. The implication,
therefore, is
that the star was already super-Eddington in the decades leading up to the giant outburst.
If Object 7 had an initial mass of $\sim 55$--$85\ M_{\odot}$ (see below),  then
its present-day mass would
be $\sim\ 25$--$40\ M_{\odot}$ \citep[from the models by][]{meynet03} 
and, at its inferred present-day luminosity, 
$\sim 10^{5.8-6.2}\ L_{\odot}$ (see below), 
the star may be currently just at the Eddington limit for its mass.

We analyzed the STIS spectrum of Object 7, shown in Figure~\ref{figspec}, by fitting the profile of 
the H$\alpha$ emission line with a pair of Lorentz profiles, one broad and one narrow, 
using the routine {\sc LMFIT} within 
IDL\footnote{IDL is the Interactive Data Language, a product of Research Systems, Inc.}. The 
central wavelengths, $\lambda_{\rm cen}$, for each of the profile components were allowed to 
roam during the fit. The resulting parameters are $\lambda_{\rm cen}=6580.78 \pm 0.80$ \AA\ and
$\gamma=6.83 \pm 1.45$ \AA\ for the broad component, and 
$\lambda_{\rm cen}=6579.19 \pm 0.04$ \AA\ and $\gamma=0.40 \pm 0.04$ \AA\ for the narrow 
component. (The term $\gamma$ is the HWHM scale parameter for the function.) The model fit is
also shown in the figure. The velocity width of the broad component is then 
$\sim$311 km s$^{-1}$. The overall profile is asymmetric, as reflected in the slightly
different $\lambda_{\rm cen}$ for the two components of the fit, 
with the profile appearing 
somewhat broader toward the red wing of the line than toward the blue wing. 

\bibpunct[ ]{(}{)}{;}{a}{}{;}
The uncertainties in the 2008 F555W and F658N magnitudes for Object 7 are appreciable, and, 
unfortunately, 
no other recent broadband {\sl HST\/} images exist to provide us with direct color information
for the star (the SN 1961V environment is not detected in the F450W and F814W
observations from 2007 by program GO-11119, PI: S.~Van Dyk). Nonetheless,
we attempt to produce a physically 
plausible and reasonably self-consistent model for Object 7 as an extinguished luminous blue 
emission-line star. We included the model emission line together with the model spectra 
of blue supergiants at LMC metallicity 
\citep[][consider the SN environment to be subsolar]{kochanek11}.
\bibpunct[, ]{(}{)}{;}{a}{}{;}
The model supergiants were those from \citet{kurucz93} within the 
STSDAS routine SYNPHOT in IRAF. 
Since the star has apparently declined in brightness only slightly over the years, we make the
assumption that the star's colors (in the WF/PC-1 bandpasses) have not varied from those 
measured in 1991, i.e., 
$m_{\rm F555W} - m_{\rm F702W} = 0.91 \pm 0.09$, 
$m_{\rm F555W} - m_{\rm F785LP} = 0.50 \pm 0.18$, and
$m_{\rm F702W} - m_{\rm F785LP} = -0.41 \pm 0.18$ mag \citep[see][their Table 4]{filippenko95}.
The coolest effective temperature (spectral type) the star could have, and still be responsible for 
photoionizing a putative surrounding nebula from which the H$\alpha$ emission arises, is
$T_{\rm eff} = 30000$ K. 
We allowed the model to be as hot as $T_{\rm eff} = 40000$ K, approximately the highest
effective temperature that supergiants in the LMC can attain \citep{massey09}.

We allowed the extinction to the model star also to vary, assumed
a \citet*{cardelli89} Galactic reddening law, corrected the model to the redshift of the host
galaxy, and normalized the model to the $m_{\rm F555W}$ observed in 2008, all within 
SYNPHOT. In order for the resulting photometry from the model star to be constrained by the 
uncertainties in the 
WF/PC-1 colors, the assumed extinction could only vary from $A_V\simeq 1.8$ to 2.3 mag. One
of the most important aspects of the model is that we can synthesize the observed F658N 
magnitude, without any further normalization, {\em exactly}. This is irrespective of the assumed 
temperature or extinction for the 
model, since the H$\alpha$ emission line dominates the model's overall spectral energy 
distribution (SED).

\bibpunct[; ]{(}{)}{;}{a}{}{;}
The synthetic $V$ magnitude of the model for Object 7, for the range of temperature and 
extinction, is 24.77. For the range in extinction to the star and the distance to the host galaxy, 
this is a physically realistic 
$M^0_V \simeq -6.9$ to $-7.4$ mag. 
\citep[Note that this is quite comparable to the $M_V \simeq -7.4$ mag brightness for the 
precursor of the modern impostor prototype, SN 1997bs in M66;][]{vandyk00}.
Integrating over the flux in the star models, after correcting for extinction and the distance to
the host galaxy, we find that the bolometric luminosity is 
$L_{\rm bol} \simeq 10^{(5.8-6.2)} L_{\odot}$. 
This range corresponds to bolometric magnitudes
of $M_{\rm bol} \simeq -9.8$ to $-10.8$ (assuming $M_{\rm bol} (\sun)$ = 4.74 mag).
The inferred range in bolometric corrections, ${\rm BC}_V \simeq -2.9$ to $-3.9$ mag, is
comparable to the ${\rm BC}_V$ found for O-type supergiants at LMC metallicity \citep{massey09}.
We show the range in the hypothetical properties for the model star in Figure~\ref{fighrd}. 
In the figure we also show 
massive stellar evolutionary tracks from \citet{lejeune01} for
subsolar metallicity ($Z=0.008$) at 120, 85, and 60 $M_{\odot}$. It
appears that the locus of Object 7 is of a star with initial mass $\sim 55$--$85\ M_{\odot}$ near the 
point in the track where the star would evolve to the WR phase.
Note that our model star is 
nowhere near as luminous, or as massive, as $\eta$ Car \citep[][cf.~their Figure 9]{humphreys94}. 

The observed F658N brightness corresponds to an integrated flux in the model H$\alpha$ 
emission line, 
$F_{{\rm H}\alpha} \simeq 6.3 \times 10^{-16}$ erg cm$^{-2}$ s$^{-1}$.
Correcting for the host galaxy distance and the range of possible extinction, and also assuming
$A_{{\rm H}\alpha}=0.81\ A_V$ \citep[e.g.,][]{parker92}, the H$\alpha$ luminosity is in the range of
$L^0_{{\rm H}\alpha} \simeq 2.5$ -- $3.6 \times 10^{37}$ erg s$^{-1}$.
This is roughly comparable to the quiescent, although variable, luminosity 
$L^0_{{\rm H}\alpha} \approx 1.4 \times 10^{37}$ erg s$^{-1}$, for $\eta$ Car A 
\citep[][their Figure 5 --- we have assumed the flux from
that figure near maximum, i.e., $\sim 6 \times 10^{-9}$ erg cm$^{-2}$ s$^{-1}$]{martin04}, 
assuming the distance to $\eta$ Car of 2.3 kpc and $A_V \sim 1.7$ mag to the central star 
\citep{davidson97}. 

The ionizing fluxes (in cm$^{-2}$ s$^{-1}$) from our model star would be $\log q_0\simeq 23.5$ for
$T_{\rm eff}=30000$ K and $\log q_0\simeq 24.3$ for $T_{\rm eff}=40000$ K \citep{martins05}.
From the bolometric luminosity, above, the star would have a radius of $R \simeq 34\ R_{\odot}$
for 30000 K and $\simeq 23\ R_{\odot}$ for 40000 K.
The number of Lyman continuum photons is then in the range of 
$\sim 2.2$--$6.6 \times 10^{49}$ s$^{-1}$.
For Case B recombination, assuming an electron density $n_e \sim 10^2$ cm$^{-3}$ and 
$\alpha_{\rm B}\simeq 3 \times 10^{-13}$ cm$^3$ s$^{-1}$, the Str\"omgren
sphere for the star would be in the range $\simeq 3.9$--5.6 pc. 
If the putative nebula around the star had been ejected at 2000 km s$^{-1}$ 
\citep[][from a trace of the H$\alpha$ profile of the 1962 
spectrum shown in Zwicky's Figure 2, we confirm that the FWHM of the line is 
$\sim 4675$ km s$^{-1}$, indicating that 
$v_{\rm exp} \sim 1986$ km s$^{-1}$]{zwicky64,branch71} in an eruption that began
at the end of 1960, 
then by 2008 August (the date of the most recent {\sl HST\/}
observations) the nebula would have expanded to $R_{\rm neb} \simeq 3.0 \times 10^{17}$ cm, 
or $\simeq 0.1$ pc. 
The nebula would therefore be easily photoionized. 
If we consider $R_{\rm neb}$ to be the Str\"omgren sphere radius, then the density of the nebula 
is probably more like $n_e \simeq 2.5$--$4.4 \times 10^4$ cm$^{-3}$.
We note that this is several orders-of-magnitude less dense than the $n_e > 10^7$ cm$^{-3}$ 
limit placed on SN 1961V's ejecta during outburst by 
\citet[][based on the lack of observed {[Fe\,{\sc ii}]} lines]{branch71}, implying
that the density has, naturally, declined as the ejecta have expanded over the decades. 
We also note that, assuming $n_e$, above, and a more likely value for $R_{\rm neb}$ (which we 
estimate below), we find that $L^0_{{\rm H}\alpha} \simeq 6.7 \times 10^{37}$ erg s$^{-1}$, which 
is
comparable to, yet about a factor of two larger than, what is observed; however, this also assumes 
a 100\% recombination efficiency. The fact that the two estimates for the luminosity from the 
nebula are close lends credence to the combination of our estimates for the nebular 
radius and density.

We can use the relation between the ionized gas mass in the nebula and 
the (extinction-corrected) H$\alpha$ flux from \citet[][who adapted this
from \citealt{pottasch80}, his Equation 5]{nota92}, 
\begin{eqnarray}
M_{\rm neb} (M_{\odot})=3.87 \cdot (F_{{\rm H}\alpha} [10^{-11} {\rm erg\ cm}^{-2} {\rm s}^{-1}]) \times \nonumber\\ (D [{\rm kpc}^2])^2  (T_e [10^4\ {\rm K}])^{0.85}  (n_e [{\rm cm}^{-3}])^{-1} 
\end{eqnarray}
\noindent and assuming an electron temperature $T_e\simeq 10^4$ K, and  
the ranges in both $n_e$ and extinction, above, 
this results in $M_{\rm neb}\approx 8$--$12\ M_{\odot}$.
If this nebula has been expanding at $v_{\rm exp}=2000$ km s$^{-1}$, 
the resulting energy
would be $E_{\rm kin}=\onehalf M_{\rm neb} v_{\rm exp}^2 \approx 10^{50.6}$ erg. Assuming
the conversion efficiency from kinetic to radiative energy is 10\%, this can straightforwardly 
account for
the total energy, $\approx 10^{49.6}$ erg, that Humphreys et al.~\citeyearpar{humphreys99} 
estimated both $\eta$ Car and SN 1961V had radiated during their mutual eruptions. 

In their analysis, 
\citet{kochanek11} produced model SEDs for the mid-IR emission from a putative eruption
survivor using DUSTY \citep{ivezic97,ivezic99,elitzur01}, assuming graphitic
and silicate dust with properties from \citet{draine84}. \citeauthor{kochanek11} 
assumed a spherical shell with
an inner radius of $\simeq 1.3$--$1.5 \times 10^{17}$ cm (which we show, below, is 
nearly a factor of
two too small), and normalized the emission from the DUSTY models to a survivor brightness
of $V=24$ mag (which we have shown, above, is nearly a magnitude too bright). Their 
preferred scenario included an input source for the models having a temperature of 7500 K, 
which agrees with the colors for the object in 1954 measured by \citet{utrobin87}. 
The luminosity for the illuminating source at the shell's center, 
$L_{\star} \simeq 10^{6.9}\ L_{\odot}$, is, then, nearly an order-of-magnitude more luminous
than what we have found for Object 7. These assumptions made by 
\citeauthor{kochanek11}~result in a dust shell with $A_V \sim 6$--11 mag, which
is considerably higher than even the estimates made by 
\citet{goodrich89} and \citet{filippenko95} of $\sim 4$--5 mag, let alone the far more modest 
$A_V\simeq 1.8$--2.3 mag that we contend, above, is more likely for the survivor.

\citet{kochanek11} further argue that, even if one were to assume a model more similar to the
quiescent LBV model which \citet{goodrich89} assumed (and to what we find likely to be the 
case for Object 7),
i.e., $T_{\star}=30000$ K, $L_{\star} \approx 10^6\ L_{\sun}$, and 
$A_V \simeq 2$ mag (corresponding to optical depths $\tau_V=2.5$ and 4.5 
for graphite and silicate dust, respectively, input into DUSTY), 
the resulting
mid-IR emission would still be comparable to the emission from $\eta$ Car.
That is, it would still be much higher in luminosity than the constraints 
imposed by the {\sl Spitzer\/} data would allow (see their Figure 6).
Again, these models were also normalized at nearly a magnitude brighter at $V$
than what we have found for Object 7.

In spite of possibly inappropriate assumptions for the dust modeling, 
no reason exists in the first place to expect that SN impostors should 
generally be as dusty and as luminous in the mid-IR as $\eta$ Car.
In Figure~\ref{figmir} we show 
the mid-IR fluxes we have have measured from {\sl Spitzer\/} archival data 
and presented in \citet{vandyk11} for several SN 
impostors: SN 1954J/Variable 12 in NGC 2403, SN 1997bs, SN 1999bw in NGC 3198, and SN 
2000ch in NGC 3432.
We have adjusted the 
fluxes for all of these SN impostors to the distance of SN 1961V. 
For SN 1954J we have measurements from 2004 
at 3.6 and 4.5 $\mu$m, but the flux in these bands may result from cool
supergiants in that impostor's immediate environment \citep[see][]{vandyk05b}; additionally, 
we obtain only upper 
limits to the emission at 5.8--24 $\mu$m. SN 1997bs is detected in 2004 only at 3.6 and 4.5 
$\mu$m and is too confused in the longer-wavelength {\sl Spitzer\/} bands. SN 1999bw was quite 
luminous in the mid-IR in 2004 (as shown in Figure~\ref{figmir}), 
but subsequently faded in 2005 and 2006 
\citep[see also][]{sugerman04}. SN 2000ch may still 
have been in outburst when the {\sl Spitzer\/} observations of its host galaxy were
made in 2007 and 2008 \citep{pastorello10}, 
so a direct comparison with the quiescent SN 1961V is not warranted. 
Although none of these SN impostors was as luminous during eruption
as SN 1961V \citep[we note that the impostors are a diverse group of objects; 
e.g.,][]{vandyk11,smith11}, 
one can see that the SEDs, or limits to the SEDs, for all of these impostors
are less luminous in the mid-IR than $\eta$ Car (we show in Figure~\ref{figmir} 
the SED of $\eta$ Car, as detected by \citeauthor{morris99} \citeyear{morris99} with
the {\sl Infrared Space Observatory}, scaled to the distance of SN 1961V). 
In particular, 
the detections and upper limits to the emission at the site of SN 1954J, which occurred nearly
a decade before SN 1961V, are entirely consistent with the upper limits for SN 1961V. 
The oldest of the known SN impostors, $\eta$ Car, may simply be an ``oddball,'' in terms of the
dust content of its Homunculus and its luminous mid-IR emission.

Furthermore, as we have already alluded to, 
we believe that the dust modeling by \citet{kochanek11} does not include the full range of 
complicating, but realistic, components. In 
particular, applying the interstellar graphite and silicate dust properties from 
\citet{draine84} may not be appropriate for a LBV nebula. 
Evidence exists that the dust in these
shells may be dominated by amorphous silicates, which we would expect for O-rich LBV nebulae
enhanced in CNO-burning products
\citep[][see also \citeauthor{lamers96}~\citeyear{lamers96}]{umana09}.
We have therefore created two dust shell models, also using DUSTY,  
one with Fe-free Mg$_2$SiO$_4$ (forsterite) and the other with the olivine MgFeSiO$_4$,
which span essentially the range in expected optical depth for amorphous silicates in the 
wavelength range of interest.
We have input to DUSTY the optical constants from \citet{jaeger03} for the former and
\citet{jaeger94} and \citet{dorschner95} for the latter. 
If we assume material densities $\rho =3.3$ g cm$^{-3}$ for Mg$_2$SiO$_4$ and 3.7 g cm$^{-3}$ 
for MgFeSiO$_4$, we find opacities at $V$ of
$\kappa_V \simeq 7$ cm$^2$ g$^{-1}$ for Mg$_2$SiO$_4$ and $\simeq 3032$ cm$^2$ 
g$^{-1}$ for MgFeSiO$_4$ (with $\kappa_V = 3 Q_{\rm abs}/4 \rho a$, where $Q_{\rm abs}$ is
the dust absorption coefficient and $a$ is the grain radius).

Additionally, if we infer from the SN 1961V light curve that the 
outburst ended sometime in 1962, then the expansion at 2000 km s$^{-1}$ 
from this end date to the mid-date of the 
{\sl Spitzer\/} observations, i.e., early 2006, leads to 
an inner radius for the
dust sphere of $R_{\rm in} \simeq 2.7 \times 10^{17}$ cm. If we also assume that
the outburst commenced, roughly, at the end of 1960, then it effectively lasted for $\sim$2 yr,
and the shell thickness $\Delta R \simeq 1.3 \times 10^{16}$ cm, or 
${\Delta R}/R_{\rm in} \simeq 0.05$. (As \citeauthor{weis03} \citeyear{weis03} has shown, 
the thickness of observed LBV nebulae, e.g, in the LMC, is generally $>$10\%; yet, as 
\citeauthor{kochanek11} \citeyear{kochanek11} point out, the DUSTY models are 
relatively insensitive at these radii to the choice of shell thickness, and we find little difference in
the results if we set ${\Delta R}/R_{\rm in} > 0.05$.)
If the mass of the gaseous nebular shell is $M_{\rm shell} \simeq 10\ M_{\odot}$ (see
above), then, assuming a dust-to-gas ratio of 1/100, the shell's dust mass is
$M_{\rm dust} \simeq 0.1\ M_{\odot}$, which, for 
optical depth $\tau_V = 3M_{\rm dust}\kappa_V/4{\pi}R^2_{\rm shell}$ and
$R_{\rm shell} = R_{\rm in}$, 
results in $\tau_V \simeq 0.004$ and 
$\simeq 0.6$, respectively (which correspond in DUSTY to 0.133 and 1.083, respectively). 
Finally, we assume for the DUSTY modeling 
a \citet{mathis77} dust grain size distribution (with index $q=3.5$) and
that the density distribution in the spherical shell around the central source varies 
$\propto r^{-2}$.
If we represent Object 7 in DUSTY as a $T_{\rm eff} = 30000$ K blackbody with luminosity
$L \approx 10^6\ L_{\odot}$, then the dust temperatures at the inner and outer
edges of the shell are approximately 
149 and 127 K, 
respectively, for the Mg$_2$SiO$_4$ model, 
and approximately 
196 and 115 K, 
respectively, for the MgFeSiO$_4$ model.

Clearly, the values of $\tau_V$ for amorphous silicates in the shell are 
significantly smaller than those
stemming from our assumed range in extinction, $A_V=1.8$--2.3 mag (and are far smaller than
what \citeauthor{kochanek11} \citeyear{kochanek11} assumed 
for their primary scenario). 
As we have already emphasized, above, it is not necessary to 
presume that any nebula around SN 1961V is particularly dusty. For this reason, 
we further assume that the additional optical depth arises along the line-of-sight to 
SN 1961V within the host galaxy, and, therefore, we 
apply the ``LMC average'' extinction law from \citet{weingartner01}.

In Figure~\ref{figmir} we show our dust models, relative to the observed flux at $V$ for Object 7
and the upper limits on the mid-IR emission from a dust shell.
The prediction of the emission from the lower-optical-depth Mg$_2$SiO$_4$ dust model is
comfortably within the observed upper limits, while, admittedly, the MgFeSiO$_4$ model predicts
that SN 1961V should have been detected at 24 $\mu$m, although it would escape detection in 
all of the IRAC bands (whereas all of the dust models by 
\citeauthor{kochanek11}~\citeyear{kochanek11} exceed the observed limits, 
at least at 8 $\mu$m, if not in all of the IRAC bands). 
The situation for this latter model is further aggravated, of course, if we assume that Object 
7 has $T_{\rm eff}=40000$ K, increasing the overall mid-IR luminosity from the dust shell.
Regardless of the star's effective temperature, one could sidestep the lack of detected 24 
$\mu$m emission, however, by 
presuming, for instance, that the amorphous silicates in the shell are less Fe-rich, which is
certainly possible. Another 
possibility is that the dust mass in the shell is less than what we have assumed, lowering the
overall luminosity of the emission.
Furthermore,
the geometry of the ejected matter from Object 7 may be aspherical, asymmetric, or both 
\citep[e.g., $\sim$50\% of known LBV nebulae appear to show a bipolar structure;][]{weis01}.
In any event, we have shown, using both observations of other SN impostors and
what we consider to be a 
more realistic model of any dusty ejecta around the star,
that the lack of mid-IR emission from SN 1961V alone, particularly 
across all of the {\sl Spitzer\/} bands, is not
a valid argument to eliminate the possibility altogether that it is a SN impostor.

\section{Conclusions}

Based on our analysis of the available data, we consider 
SN 1961V (still) to be 
a SN impostor and that Object 7 is the survivor of this event. All of the arguments we have 
made above, as part of this analysis, support this conclusion. We have also attempted to 
dispel several erroneous suppositions about the nature of this event. 
The survivor is clearly observed in very recent years to exist. The star quite plausibly has the 
properties of a quiescent, massive LBV. 
The star has not yet exploded and is not a radio SN.  The star is surrounded by a 
circumstellar shell or nebula, however, this shell is not nearly as dusty as required by 
\citet{kochanek11} and previous investigators. We find that the visual extinction to Object 7 is
in the range $A_V=1.8$--2.3 mag, and we suggest that most of this extinction is arising from the
interstellar medium along our line-of-sight within the host galaxy, rather than from the shell itself.
As a result, therefore, the shell need not be nearly as luminous an IR
emitter as $\eta$ Car. This conclusion is further supported by the mid-IR observations of other SN
impostors.

The positional proximity of Object 7 to SN 1961V, and the positional offset
between SN 1961V and the radio source centroid, are both inescapable facts. The only  
conceivable way to counter the former fact is to conclude that Object 7 is 
either a physical or an optical double to the progenitor of an actual SN. This, of course, is
possible. The uncertainty alone in the absolute position, $0{\farcs}1$, corresponds to $\sim 5$ pc, 
certainly allowing for the possibility, in the relatively crowded cluster environment of SN 1961V, 
that Object 7 is merely a neighbor to the progenitor.
Also, the probability of core-collapse SNe for high-mass stars in binaries 
is relatively high \citep[e.g.,][]{kochanek09}. 
Although, if Object 7 {\em were\/} the binary companion to the star that exploded, we might
expect Object 7 to have been stripped or otherwise affected by what would have been 
a very powerful explosion, potentially leading to an unusual brightness or color. This is not 
supported by the available 
photometry or inferred luminosity of Object 7. Instead, it appears to have the properties of a 
``run-of-the-mill,'' evolved, high-mass star.
Nonetheless, the positional offset between SN 1961V and the radio
centroid is insurmountable --- these two are not one and the same, which is essentially at the
heart of the case that has been made for SN 1961V being a true SN.

No need exists for the supposed core-collapse explosion ``hybrid'' of a 
SN II-P and SN IIn. Furthermore, 
we disagree with the statement made by \citet{smith11} that  SN 1961V is somehow unique 
among the impostors. 
It has direct analogs in both SN 2000ch \citep{wagner04,pastorello10} 
and SN 2009ip in NGC 7259 \citep{smith10,foley11}, specifically in terms of the
inferred expansion velocities at eruption and of the light curve behavior. The recent, more 
powerful revival of SN 2000ch
displays ``unusually'' high velocities (FWHM) of 1500--2800 km s$^{-1}$ \citep{pastorello10}. 
SN 2009ip shows 
blueshifted He\,{\sc i} absorption at 3000--5000 km s$^{-1}$, which \citet{smith10} speculate 
for SN 2009ip could be due to a fast blast wave, which occurred quasi-contemporaneously with 
the origin of the slower ejecta. SN 2000ch exhibits 
multiple P Cygni absorption components, with profile edges up to 3000--3500 km s$^{-1}$. A 
sustained high-luminosity pre-eruption state was seen for both SNe 2000ch and 2009ip; 
SN 2000ch was at
$M_R\simeq -10.7$ mag prior to the 2000 eruption \citep{wagner04}, and SN 2009ip had a 
pre-eruption luminosity of $M_V\simeq -10$ mag \citep{smith10}. Although not nearly the
$M_{\rm pg}\approx -12$ mag brightness of SN 1961V, both SNe 2000ch and 2009ip,
therefore, both may also 
have been in a super-Eddington phase prior to their giant eruptions.
Even though the line widths were different, it is interesting to note that the H$\alpha$ line profiles, 
in particular, of SN 1961V \citep[][see his Figure 2]{zwicky64} and 
SN 2000ch \citep[][see their Figure 7]{wagner04}, 
near maximum of the 2000 eruption, were very similar --- a sharp drop-off to the blue wing and 
extended wing to the red. The H$\alpha$ emission line in 2002 for Object 7/SN 1961V also 
showed a similar asymmetric profile (and, we emphasize that this profile had a very similar 
shape as the much broader line profile for SN 1961V in 1962, as shown by 
\citeauthor{zwicky64}). A possible explanation could be the effects of dust extinction in the 
expanding 
ejecta, although we have shown that the extinction from the ejecta is likely relatively modest for
SN 1961V. A detailed analysis of this effect should be explored, although we consider this to be 
beyond the scope of this paper.

We do agree with \citet{smith11} and \citet{kochanek11}, however, that the ``undulations'' in SN 
1961V's post-maximum 
light curve may well have arisen from the blast wave overtaking previously-ejected shells of 
matter ahead of the 
shock. We speculate here that the star was already in a sustained, eruptive outburst prior to 1960 
and that the onset of the super-outburst, which peaked in luminosity in 1961 December, could 
possibly have been due to the interaction of the fast-moving ($\sim$2000 km s$^{-1}$), 
dense, massive shell with the pre-existing, slower, less dense mass loss. A potential
analog is the behavior of the SN IIn 1994W in NGC 4041 --- \citet{dessart09} modeled the high 
luminosity, sustained plateau, and sudden drop-off of this SN as the
interaction of a fast ($\gtrsim 1500$ km s$^{-1}$), dense shell interacting with a slower, less dense 
shell. Furthermore, both the brief and more sustained plateaus in the SN 1961V light curve could 
have arisen from the interaction of the fast blast wave with various regimes of previously ejected 
matter, or to periods of partial recombination in the expanding ejecta, analogous to 
the recombination wave in SN II-P ejecta. Alternatively, these plateaus could also have been due 
to subsequent lesser eruptions by the star (Humphreys et al.~\citeyear{humphreys99}).
The fast-moving massive shell radiated for many years after the super-outburst, albeit 
more faintly. \citet{goodrich89} detected the broad emission-line 
component at $F_{{\rm H}\alpha} \simeq 2.2 \times 10^{-16}$ erg cm$^{-2}$ s$^{-1}$ in 1986. 
We can set a limit on detection of this component in the {\sl HST}/STIS spectrum from 2002 at 
$F_{{\rm H}\alpha} < 3 \times 10^{-16}$ erg cm$^{-2}$ s$^{-1}$ (see Figure~\ref{figspec}).

\bibpunct[; ]{(}{)}{,}{a}{}{;}
It is correct to point out the similarities in the variety of SN impostor and SN IIn properties. The
two are very likely intimately intertwined, as we now know that at least one SN IIn arose from a SN 
impostor \citep[SN 2005gl;][]{galyam07,galyam09}. The SN 
IIn are also a diverse class of objects (if the term ``class'' is truly applicable in the case of such 
heterogeneity), and it is still not clear how many SNe IIn have, in fact, really been impostors. 
For instance, SN 1994W itself may be an impostor \citep{dessart09}.

Obviously, we have extrapolated fairly extravagantly from two photometric measurements of 
Object 7 from
recent {\sl HST\/} data. We have had to assume that the colors for the star from 1991 are 
applicable
to 2008 as well. What is ultimately required is a set of deep, multiband optical and near-IR
observations of the SN 1961V environment using the modern {\sl HST} with the Wide Field 
Camera 3, which would vastly 
improve upon the pre-furbishment WF/PC-1 images and the motley assortment of WFPC2 and
unfiltered STIS image data analyzed to date. Such future observations would only require a rather
modest investment of {\sl HST\/} time --- to image in $BVRIJH$ 
at $S/N\gtrsim 10$
would require only 3 orbits. Looking beyond {\sl HST}, 
observations with the {\sl James Webb Space Telescope\/}, particularly with the Mid-IR Instrument
(MIRI), would reveal the actual 
dust emission from the surviving star. 
This object has clearly garnered much attention
and speculation over the decades and continues to provide us with invaluable insights into the
evolution of the most massive stars. With these space-based data, the definitive nature of the SN 
1961V precursor may well be established once and for all. 

\bibpunct[; ]{(}{)}{;}{a}{}{;}
Finally, NGC 1058 should continue to be regularly monitored by SN searches in 
nearby galaxies. 
For Object 7 as the SN 1961V survivor, we might expect, 
much as was the case for SN 2006jc \citep{foley07,pastorello07}, that the star will {\em truly\/} 
explode as the SN that we believe other authors have erroneously concluded has already 
occurred. The super-outburst from the 1960s could well be the forerunner of a core-collapse 
event. It is possible that the star will explode as a high-luminosity, hydrogen-rich SN IIn, 
such as SN 2006gy \citep{ofek07,smith07b}, or as a more helium-rich SN Ibn, such as SN 2006jc 
\citep[as suggested by][]{kochanek11}.

\acknowledgments 
This work is based in part on archival data obtained with the {\sl Spitzer Space Telescope}, which 
is operated by the Jet Propulsion Laboratory, California Institute of Technology under a contract 
with NASA. 
We thank Roberta Humphreys, for her comments, and the referee, whose 
recommendations helped improve the manuscript. 
We also thank Chris Kochanek for correcting us on our use of ``$\tau$'' in the DUSTY modeling.

{\it Facilities:} \facility{VLA}, \facility{HST}, \facility{Spitzer}.

\clearpage

\clearpage

\begin{figure}
\figurenum{2}
\plotone{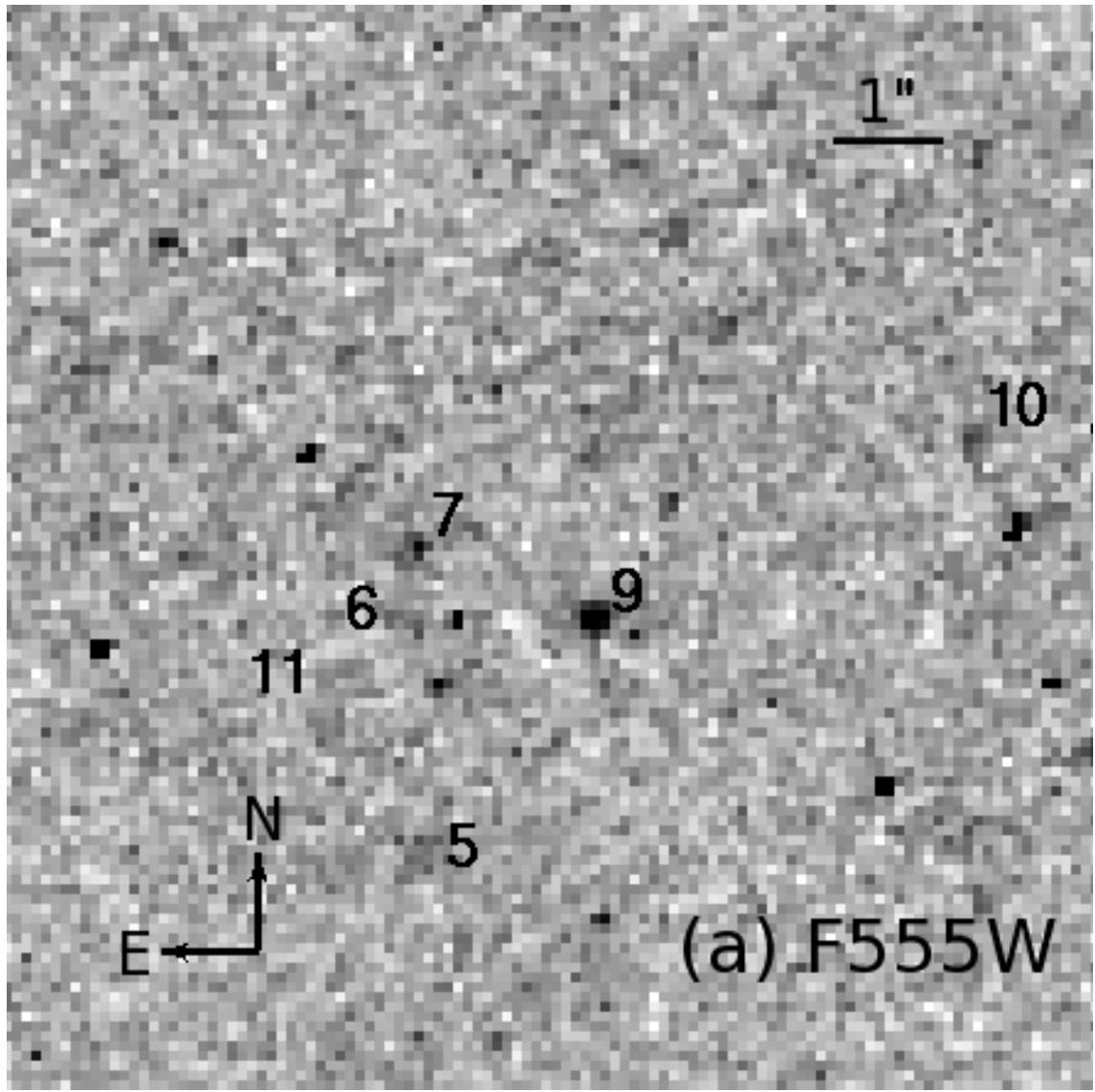}
\caption{{\sl HST\/}/WFPC2 image of the SN 1961V site from 2008, obtained from the archive, 
(a) in F555W, and (b) in F658N.
Source numbering is the convention from \citet{filippenko95} and \citet{vandyk02}.  
North is up, and east is to the left.\label{hstimage}}
\end{figure}

\clearpage

\begin{figure}
\figurenum{2}
\plotone{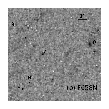}
\caption{Continued.}
\end{figure}

\clearpage

\begin{figure}
\figurenum{3}
\plotone{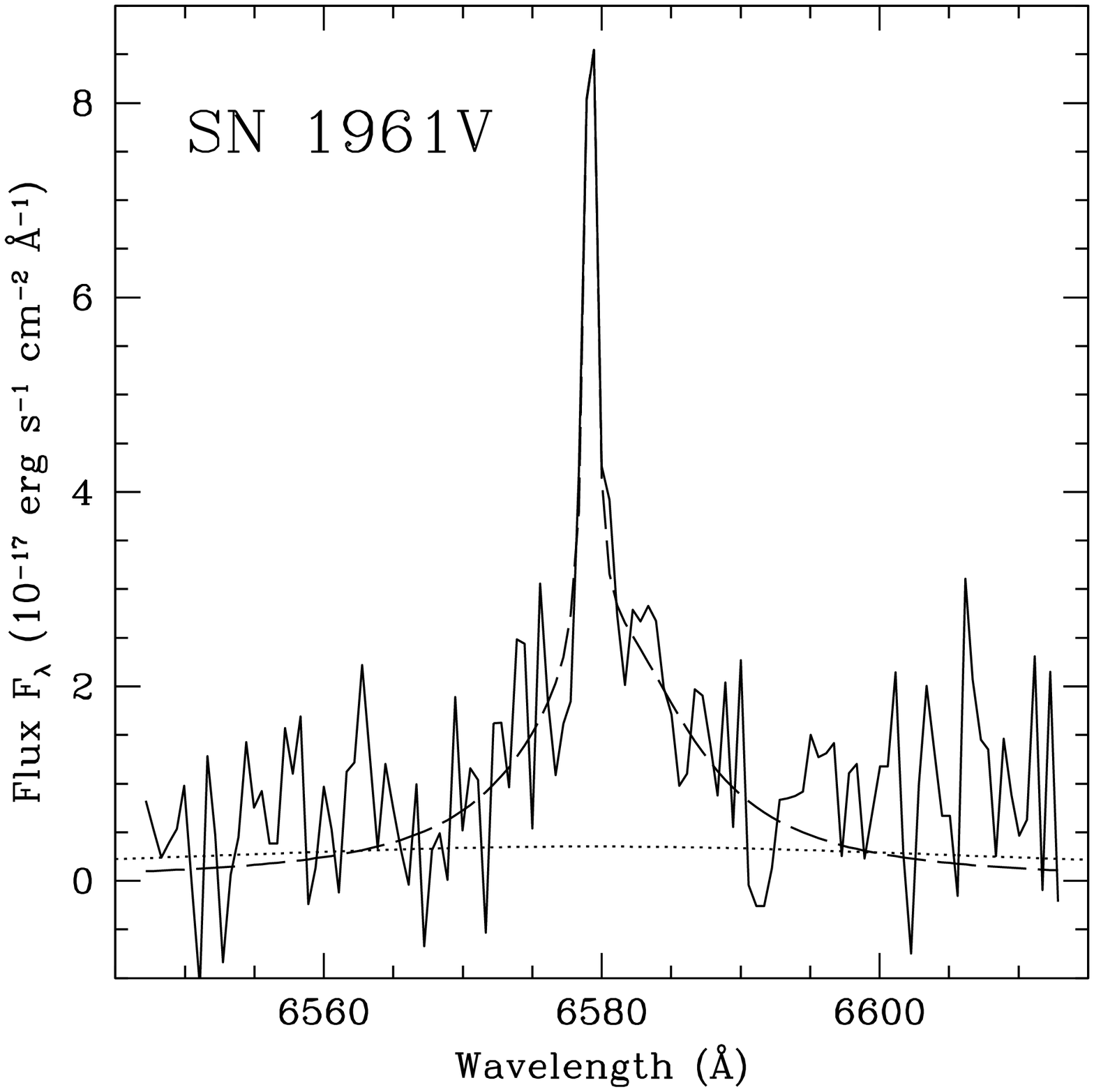}
\caption{A re-extraction of the {\sl HST}/STIS spectrum ({\it solid line}) of SN 1961V obtained by 
\citet{chu04} in 2002, showing the only distinct emission line visible in that spectrum, i.e., 
H$\alpha$. We have fit both the broad and narrow profiles of the line with the combination
of two Lorentz functions ({\it dashed line}; see text for details). We also show the limit on the
detection of the very broad ($\sim$2000 km s$^{-1}$) component to the emission line
({\it dotted line}), first observed by \citet{zwicky64} and later by \citet{goodrich89}.\label{figspec}}
\end{figure}

\clearpage

\begin{figure}
\epsscale{1.00}
\figurenum{4}
\plotone{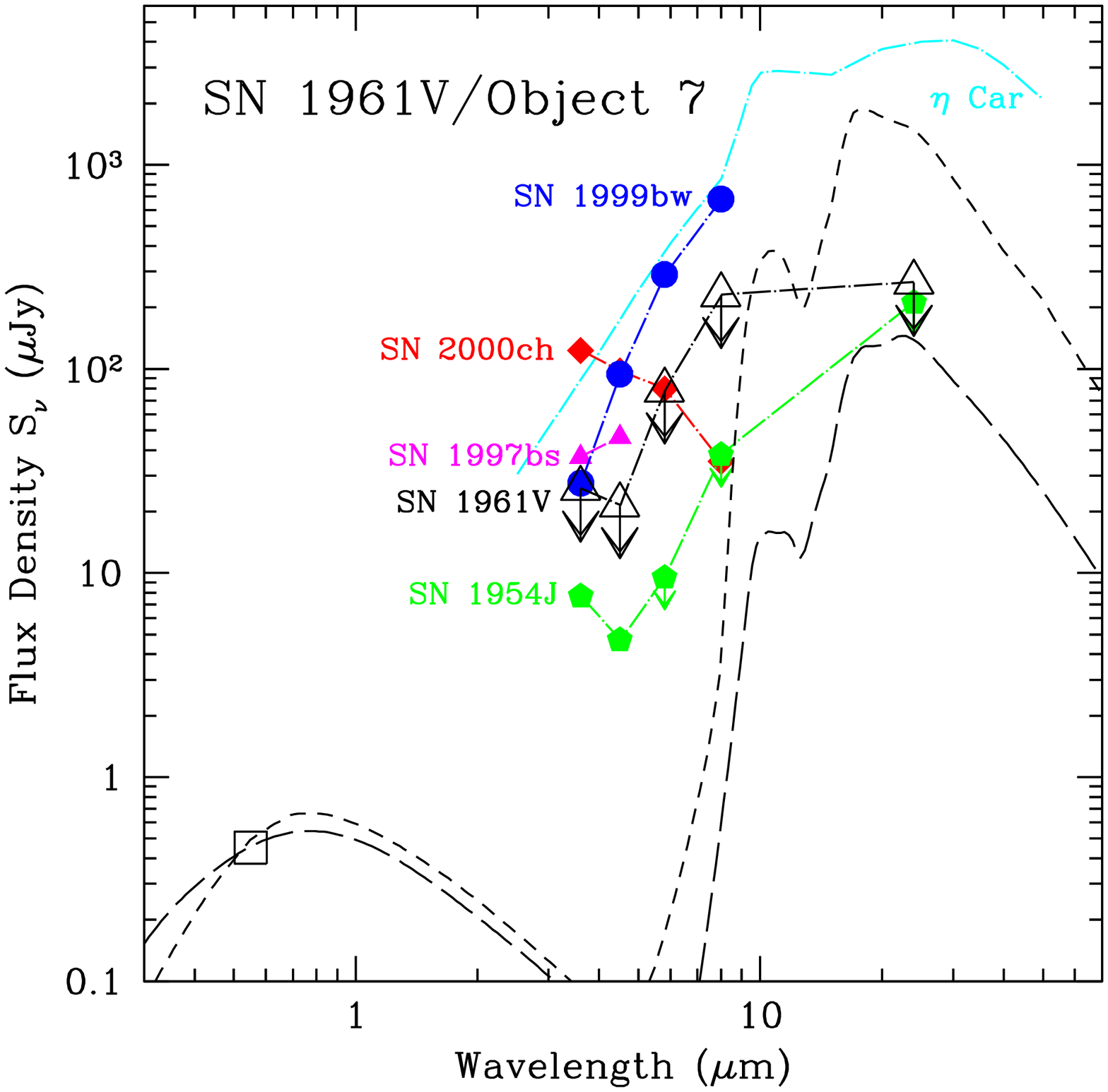}
\caption{Upper limits to the flux densities in the {\sl Spitzer\/} IRAC and MIPS 24 $\mu$m bands 
({\it open triangles}) at the exact position of SN 1961V \citep{klemola86}. Also shown are
measurements (and upper limits to detection) in some or all of these bands \citep{vandyk11} 
for the ``impostors'' SN 1954J/Variable 12 ({\it green filled pentagons}), SN 1997bs 
({\it magenta filled triangles}), SN 1999bw ({\it blue filled circles}), and SN 2000ch
({\it red filled diamonds}). Additionally, we show the observed SED for $\eta$ Car
\citep[][{\it cyan dot-dashed line}]{morris99}.
Finally, we show model SEDs generated using
DUSTY \citep{ivezic97,ivezic99,elitzur01}, 
\bibpunct[; ]{(}{)}{,}{a}{}{;}
assuming a spherical shell of inner
radius $R_{\rm in} \approx 2.7 \times 10^{17}$ cm and thickness
$\Delta R/R_{\rm in} = 0.05$, around a central
illuminating source assumed to be a $T=30000$ K blackbody with luminosity
$L \approx 10^6\ L_{\odot}$; the dust in the shell is
assumed to consist of amorphous silicates, either Mg$_2$SiO$_4$ 
\citep[][{{\it long-dashed line}}]{jaeger03} or MgFeSiO$_4$ 
\citep[][{{\it short-dashed line}}]{jaeger94,dorschner95}. 
We also assume additional extinction, following the ``LMC average'' law from
\citet{weingartner01}, in the foreground within the host galaxy along our 
line-of-sight. 
The total visual extinction for the models is assumed to be the midpoint value in the range
$A_V=1.8$--2.3 mag, and the models 
are normalized to the $V=24.77$ mag brightness of the SN 1961V survivor, Object 7 ({\it open square}). See text.\label{figmir}}
\end{figure}

\clearpage

\begin{figure}
\epsscale{1.00}
\figurenum{5}
\plotone{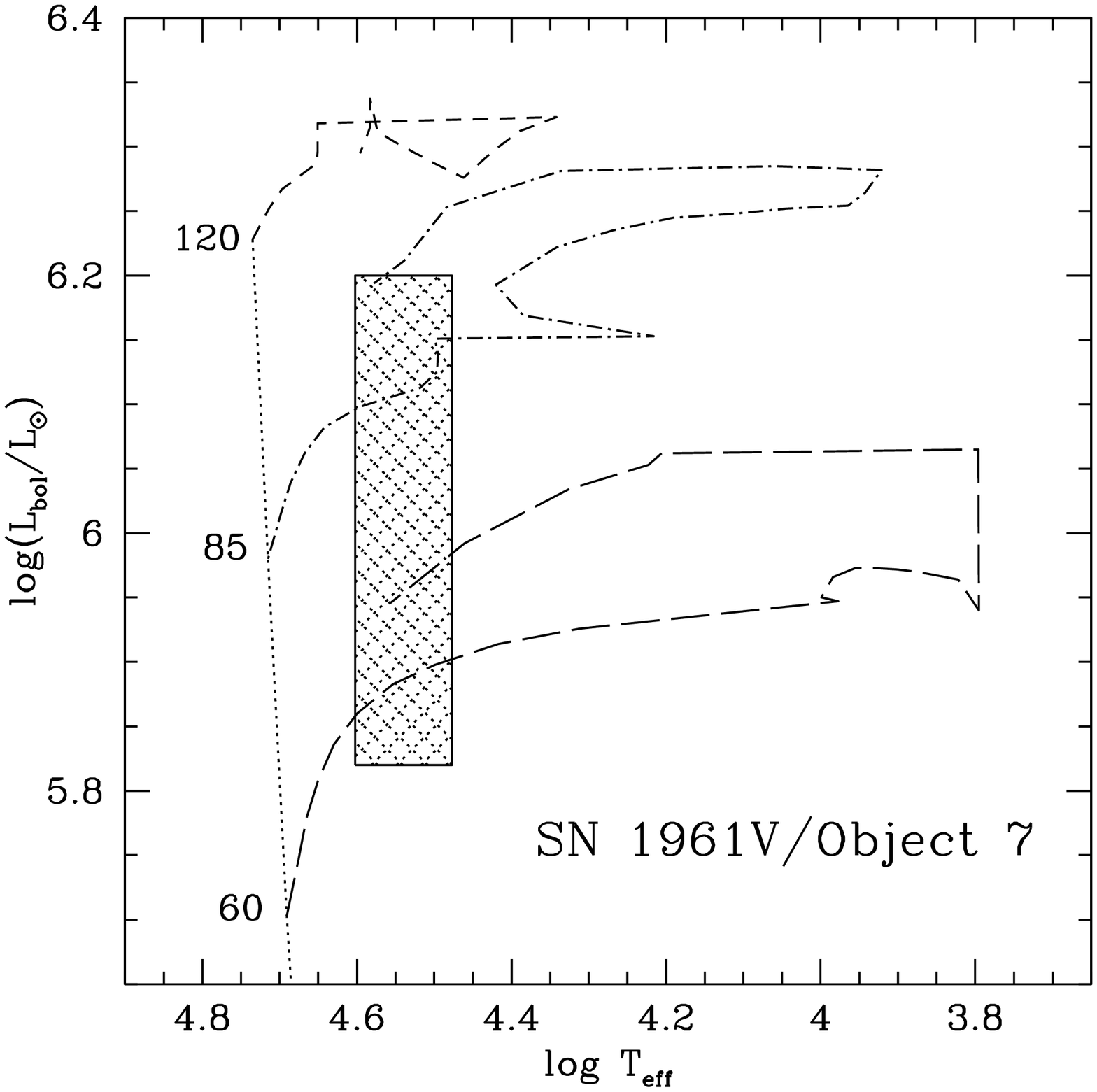}
\caption{Hertzsprung-Russell diagram, showing the approximate locus of properties for our 
hypothetical model survivor of SN 1961V at quiescence ({\it shaded region}). 
Also shown are theoretical evolutionary tracks for subsolar metallicity ($Z=0.008$) at 120 
({\it short-dashed line}), 85 ({\it dot-dashed line}), and 60 $M_{\odot}$ ({\it long-dashed line}), 
from \citet{lejeune01}. This figure intentionally emulates a similar figure in 
\citet[][their Figure 9]{humphreys94} for direct comparison.\label{fighrd}}
\end{figure}

\end{document}